\documentstyle[12pt]{article}
\newcommand{\be}{\begin{equation}}
\newcommand{\ee}{\end{equation}}
\newcommand{\s}{\section}
\newcommand{\ci}{\cite}
\newcommand{\r}{\ref}
\newcommand{\p}{\partial}
\newcommand{\eps}{\epsilon}
\newcommand{\al}{\alpha}

\newcommand{\arrow}{\rightarrow}

\begin{document}

\begin{titlepage}

\vspace{2.0cm}
\begin{center}
{\Large{\bf Two dimensional Quantum Chromodynamics
as the limit of higher dimensional theories $^*$}}\\

\vspace{1.7cm}

{\large A. Ferrando$^{\dagger}$ and A. Jaramillo$^{\dagger\dagger}$}\\

\vspace{0.2cm}
{Departament de F\'{\i}sica Te\`{o}rica and I.F.I.C.}\\
{Centre Mixt Universitat de Val\`{e}ncia -- C.S.I.C.}\\
{E-46100 Burjassot (Val\`{e}ncia), Spain.}

\vspace{2.0cm}
{\bf Abstract}
\vspace{0.2cm}
\begin{quotation}
{\small We define pure gauge $QCD$ on an infinite strip of
width $L$. Techniques similar to those used in finite $TQCD$
allow us to relate $3D$-observables to pure $QCD_2$ behaviors.
The non triviality of the $L \arrow 0$ limit is proven and the
generalization to four dimensions described. The spectrum of
the theory in the small width limit is analyzed and compared
to that of the two dimensional theory.}
\end{quotation}

\end{center}
\vspace{2.0cm}
$^*$Supported in part by CICYT grant \# AEN93-0234 and DGICYT
grant \# PB91-0119. \\
$^{\dagger}$ ferrando@evalvx.ific.uv.es\\
$^{\dagger\dagger}$ jaramillo@evalvx.ific.uv.es\\

\end{titlepage}

\s{Introduction}
Since the pioneering work of 't Hooft \ci{th74} Quantum
Chromodynamics in two dimensions ($QCD_2$) has been used as a
model to understand some of the features of the true theory. It
shares with its four-dimensional version many properties:
confinement \ci{ei76,fv94}, existence of massless pseudoscalars \ci{ac82,bl82}
and asymptotic freedom \ci{cc76}. The dimensionality of space-time
allows the unveiling of the mechanism behind these in an exact
manner. The drastic reduction in dimension, the only
approximation when considering $QCD_2$ as a model for $QCD$ , is
however too extreme.

In here we develop a continuous procedure to lower the
dimensionality of a gauge theory. We start by considering $QCD$
in $2+1$ dimension. One of the spatial dimensions is considered
of finite length $L$ ({\it the strip}) and we study how the three
dimensional theory approaches the two dimensional one as the
width of the strip is made to vanish. The choice of gauge is crucial
for the process to proceed smoothly. We pay special attention to
the spectrum and compare our findings with those of recent
developments \ci{dk94,sm94}.

The theory on the strip contains besides the conventional $QCD_2$
(2D) gluons an infinite set of covariant (non-gauge) interacting
fields $(\phi,\{V^n_{\alpha},\; \alpha = 0,1 ; \; n \neq 0 \})$
belonging to the adjoint representation of the color group.

The $\phi$ field is a massless gauge-invariant degree of freedom.
It is the remnant of the gauged-away {\it transversal} field $A_2$.
This scalar field  couples to the 2D gluons through the covariant
derivative to preserve the {\it longitudinal} gauge invariance.
It interacts with the $V$ fields and undergoes
a mass renormalization process which provides it with mass. Unlike the
2D gluons, whose masslessness is protected by the 2D gauge invariance,
there is no custodial symmetry for the scalar field. Moreover the
non-abelian scalar field influences the 2D gluon dynamics by dressing
the coupling constant $g_2$.

The $V$ fields of the non abelian strip theory  self-couple
with couplings which are not arbitrary, but determined by the 3D gauge
invariance, lost after the implementation of the gauge fixing condition.
The couplings are not gauge invariant under the 2D gauge
symmetry because these modes are massive. The 2D gauge symmetry,
however is instrumental in fixing the couplings of the 2D gluons
with the $V$ fields through covariant derivatives. All the couplings
preserve the $U(1)^n$ topological conservation law, so that the global
$n$-charge will be conserved in any vertex.

The theory on the strip carries  a natural infrared cutoff,
the strip width $L$. In the effective 2D action this scale parameter
generates the perturbative masses of the $V$ fields. These masses do not
and should not survive the infinite volume limit, since they arise from
the boundary conditions. However in $QCD$ we expect mass scales to arise
in the form of the renormalized 2D coupling constant $g_2$ by the effect
of the scalar and $V$-particle loops \footnote{Certainly the boson loops
will provide the gluon correlator with an $L$ dependence additional
to that coming from the bare masses. If and how a mass parameter
is generated depends on the resolution of the renormalization
program.}. This mass scale is fundamental in characterizing
confinement in 3D \ci{fj94b}.

\s{Non-abelian gauge theory on the strip}
A pure $SU(N)$ gauge theory defined on an infinite strip of width
$L$ has the following action

\be
 S(L) = \int_0^L \!\! d\bar{x} \int \!\! dx
                        \left\{-\frac{1}{4}
                             F_{\mu\nu}(x, \bar{x}) F_{\mu\nu}(x, \bar{x})
                                            \right\}
\label{3D}
\ee
We are calling the infinitely extended coordinates (space and
time) jointly by $x$ and the bounded transversal one by
$\bar{x}$. We compactify the transversal degrees of freedom by
requiring periodic boundary conditions on the field strength
$F_{\mu \nu}^a$. In order to describe the gauge dynamics on the
strip we need the boundary conditions on the gauge field
\be
A^a_{\mu}(x, \bar{x}+L) = A^a_{\mu}(x, \bar{x})
\label{bc}
\ee
Periodicity implies that $A^a_{\mu}$ can be expanded in Fourier
modes
\be
 A^a_{\mu}(x, \bar{x}) =  \frac{1}{\sqrt{L}} \left[a^a_{\mu} +
\sum_{n=1}^{\infty}
	\left( V^{na}_{\mu}(x)e^{i\frac{2\pi n}{L}\bar{x}} +
	       V^{na\ast}_{\mu}(x)e^{-i\frac{2\pi n}{L}\bar{x}} \right)
								\right]
\label{FourierExpansion}
\ee

Our aim is to eliminate the transversal component of the gauge
field by an appropriate selection of the gauge. This can be done
for all transverse modes except for the zero mode. In order to
avoid the appearance of ghosts we choose the gauge $\p_2 A_2 =0$ \ci{dh82}
and obtain as a gauge condition for the Fourier modes (\r{FourierExpansion}),

$$ a^a_2(x) = \phi^a(x) =  \frac{1}{L}
\int_0^L d\bar{x} A_2^a(x,\bar{x})$$
\be
V^{na}_2 = 0 \; ; \; \forall n \neq 0
\label{gf}
\ee
The gauge field zero mode is therefore a gauge invariant object
\footnote{Notice that the standard axial gauge $A_2=0$ is not allowed on the
strip because $\phi$ cannot be removed by a gauge transformation.}.
Moreover the above equations allow us to relate $a^a_2$ to a very
familiar object in Quantum Field Theory at finite temperature,
the so called Polyakov loop. On the strip we may define the
{\it transverse} Polyakov loop as
\be
Tr g(x) = Tr \exp [ (i g_3\sqrt{L}) \phi^a T_a]
\ee
where $g_3$ is the coupling constant of the 3D theory. As the
ordinary Polyakov loop $Tr g(x)$ is gauge invariant and it cannot
be suppressed by a gauge transformation because it is a physical
degree of freedom. Nevertheless the physical interpretation of
the time and transverse Polyakov loops are very different.
Therefore analogies between them beyond their formal properties
must be avoided.

Once the gauge has been chosen it is possible to write the new
action in terms of the Fourier modes of the gauge field and
then perform the integration in the transversal coordinate
obtaining in this way the 2D action
\begin{eqnarray}
 \lefteqn{{\cal L}_2 = \frac{1}{2} f_{\alpha\beta}^2
               +  (D_\alpha \phi)^2 +}
                                        \nonumber \\
     &&   \sum_{n \neq 0} \left\{
        \frac{1}{2} (D_\alpha V^n_\beta-D_\beta V^n_\alpha)
               (D_\alpha V^{n\ast}_\beta - D_\beta V^{n\ast}_\alpha)
                        + V^n_\alpha M_n^2(\phi)V^{n\ast}_\alpha
                                       \right.    \nonumber \\
	&&  \left. - i \frac{g_3}{L^\frac{1}{2}} f_{\alpha\beta}
			[V_\al^n,V_\beta^{n\ast}]  \right\}  \nonumber  \\
         &&  -i \frac{g_3}{L^\frac{1}{2}}
                \sum_{n,l \neq 0} (D_\alpha V^{n\ast}_\beta-D_\beta
					V^{n\ast}_\alpha)
                                [V^l_\alpha, V^{n-l}_\beta]   \nonumber \\
        && - \frac{g_3^2}{2L} \sum_{l,l' \neq 0}
                [V^{l\ast}_\alpha, V^{l}_\beta]
                        [V^{l'\ast}_\alpha, V^{l'}_\beta]   \nonumber \\
        &&  - \frac{g_3^2}{2L} \sum_{n,l,l' \neq 0}
                [V^{l\ast}_\alpha, V^{n+l}_\beta]
                        [V^{l'\ast}_\alpha, V^{l'-n}_\beta]
\label{2D}
\end{eqnarray}

The squared color matrix $M_n^2$ is given in terms of the scalar
Field by
\be
  M_n^2(\phi)_{ab} =
		(m_n^2 + \frac{g_3^2}{L} \phi^2) \delta_{ab}
			-\frac{g_3^2}{L} \phi_a \phi_b
                        - 2 i \frac{g_3}{L^{1/2}} m_n \epsilon_{abc} \phi_c
  \label{Mass}
\ee
The covariant derivative is defined by
\be
D_\alpha \equiv \p_\alpha -
i \frac{g_3}{L^\frac{1}{2}}[a_\alpha,.].
\ee
Finally the zero mode tensor $f_{\alpha \beta}$ is just the
ordinary 2D gluon term
\be
 f_{\alpha\beta} =
        (\p_\alpha a_\beta - \p_\beta a_\alpha)
                - i \frac{g_3}{L^\frac{1}{2}} [a_\alpha,a_\beta]
        \label{fab}
\ee
Since our final goal is to connect the theory on the strip to
two dimensional theories we must scale the coupling constant in
the proper way. Our choice of normalization for the basis of the
Fourier expansion leads to
\be
 [a_\alpha] = [V^n_\alpha] = [\phi] = 1
 \label{normal}
\ee
the familiar dimensions of the two dimensional fields.
Therefore only the coupling constant needs to be scaled in order
to recover the canonical $QCD_2$ dimension , $[g_2] = M$.
Its scaling becomes
\be
 g_2 = \frac{g_3}{L^\frac{1}{2}}
 \label{g}
\ee

The effective 2D action Eq.(\ref{2D}) has a global
$[U(1)]^n$ $(n \arrow \infty)$ symmetry whose associated
conserved charges are the mode numbers. Its quantization is
a necessary condition for the fields to fulfill the required
periodic boundary conditions Eq.(\ref{bc}).

The effective 2D action is invariant under what we call
{\it longitudinal} gauge transformations. This is the symmetry left
after imposing the gauge fixing conditions, Eqs.(\ref{gf}).
We observe that any gauge transformation behaving as
\be
\partial_2 U(x,\bar{x}) = 0 \Leftrightarrow U = U(x)
\ee
preserves the axial gauge conditions and transforms the 2D
Fourier fields as
\begin{eqnarray}
 a_\alpha & \rightarrow & U a_\alpha U^\dagger  - \frac{i}{g_2}(\p_\alpha
U)U^\dagger
                                        \nonumber \\
                  \phi & \rightarrow & U \phi U^\dagger
                                        \nonumber \\
                  V^n_\alpha & \rightarrow & U V^n_\alpha U^\dagger
 \label{2ggauge}
\end{eqnarray}

The non-covariant transformation law of the $a_{\alpha}$ field
guarantees the covariance of the 2D field strength $f_{\alpha
\beta}$ and thus the invariance of the first term of
Eq.(\ref{2D}). For this reason these fields describe the two
dimensional gluons. Besides the conventional gluons we find an
infinite set of covariant (non-gauge) interacting fields $(\Phi,
\{V^n_{\alpha}, n=1,...\}$ belonging to the adjoint representation
of the color group. Since they carry color charges they interact with
the 2D gluons in a gauge invariant way. The derivatives terms of
these latter fields are defined in terms of covariant derivatives,
the non derivative terms depend on the covariant field $V^n_{\alpha}$
and therefore the invariance of the action (\ref{2D}) under
{\it longitudinal} gauge transformations is explicit.

\s{The small $\epsilon$ regime of $QCD$ on the strip}

In order to study the 2D effective action it is convenient
to change the parameters defining the theory ($g_3, L$) into a
more befitting set to work in 2D,

\begin{eqnarray}
        g_2 \equiv \frac{g_3}{L^{1/2}}; & &[g_2] = M \\
        \epsilon \equiv g_3 L^{1/2}; & &[\epsilon] = 1
\label{2Dp}
\end{eqnarray}

At a given width $L$ and coupling constant $g_3$ we can characterize our
theory by fixing the 2D gauge coupling constant $g_2$ and the
dimensionless width $\epsilon$. As far as the effective 2D theory
is concerned $\epsilon$ acts as a dimensionless coupling constant.
We proceed to find an expansion in this coupling.

In a naive analysis, and since  the lagrangian masses of the $V$ fields are

\be
  m_n = \frac{2 \pi n}{L} = \frac{2 \pi n g_2}{\epsilon}
\label{Vmasses}
\ee
we would expect the non-zero mode fields to disappear in the $\epsilon
\arrow 0$ limit. Indeed, and just on classical grounds,  the
integration of the infinitely heavy particle sector has no consequence
on the effective action for the remaining massless degrees of freedom,
that is, the 2D gluons and the scalar field. At tree level, the sole
contribution of the non-zero modes to the reduced action
(\ref{2D}) would be through the substitution of the $V$
field by its {\em vev} at $\epsilon=0$

\be
 V^n_\al \arrow < \!V^n_\al\!> = 0
\label{Substitution}
\ee
 So that,

\be
 {\cal L}_{cl}^{QCD_3} =  \frac{1}{2} f_{\alpha\beta}^2
               +  (D_\alpha \phi)^2
\label{ClassicalQCD}
\ee
However the classical approach Eq.(\ref{ClassicalQCD}) does not provide
the correct answer to the problem of the real $\epsilon \arrow 0$
limit of $QCD$ on the strip. The reason is that there are important quantum
effects because the loop and $\epsilon$ expansions are {\em not} independent.

The non-zero mode fields become infinitely heavy
in the $\epsilon \arrow 0$ limit and a good description
of the system can be given in terms of the lighter degrees of freedom.
However the integration of the heavy modes must be performed in a more
accurate way than previously (Eq.(\ref{Substitution})) presented.
The effective potential for  2D gluons and scalar will include contributions
arising from many heavy $V$ loop diagrams. In all of them, 2D gluons and
scalar particles will appear as external legs. The dependence on $\epsilon$ of
these multi-loop diagrams will come through the $V$ masses
Eq.(\ref{Vmasses}) exclusively. Thus we can find the order in $\epsilon$ of
any diagram just by analyzing its $m_n$ dependence.

Consider an arbitrary diagram of degree C in the coupling constant
($\sim g_2^C$) and B external legs at zero momentum,
with a complicated structure in terms
of the vertices and propagators defined by the Feynman rules
of the action (\ref{2D}). Its behavior can be characterized as
\footnote{Recall that all the operators contributing to the effective
potential have to have mass dimension equals 2.}

\be
 \mbox{diagram} \sim (g_2)^C m_n^D = (g_2)^{C+D} \epsilon^{-D}
					=(g_2)^2 \epsilon^{-D}
\label{OrderDiagram}
\ee
where $D$  can be expressed in terms of the superficial degree
of divergence $d$, the number of internal transversal propagators
$i_{(VV)_\perp}$, and the number of external $\phi VV$ vertices of the
diagram by

\be
 D = d - 2i_{(VV)_\perp} + b_{\phi VV}
\label{D}
\ee
Using standard topological graph relations, we obtain

\be
 d = 4 - 2 l - b' - 2b_{\phi VV}+2i_{(VV)_\perp}
\label{d}
\ee
where $l$ is the number of loops and $b'$ stands for the number of
external legs not of the  $\phi VV$  kind. That is

\be
 b = b' + b_{\phi VV}
\label{B}
\ee
After some elementary algebra we obtain an expression relating the
$D$ to the number of loops $l$,

\be
 D = 4 - 2l - b
\label{DL}
\ee

It is now easy to understand why the classical action (\ref{ClassicalQCD})
cannot be the real $\epsilon \arrow 0$ limit of $QCD_3$. This limit is
defined by all the possible $D=0$ operators we can construct out of the
external 2D gluon and scalar fields. Since

\be
        \{ b  \geq  2 , l \geq  0 \}
\label{BL}
\ee
the above condition can be fulfilled only if

\be
       \{b = 4 , l = 0\}\;\; \mbox{or}\;\; \{b = 2 , l  = 1\}
\label{D=0Conditions}
\ee
The only operators which can appear in the effective action verifying the first
condition {\em and} compatible with gauge invariance are

\begin{eqnarray}
	f_{\al\beta}^2
	& \stackrel{a_\al = u_\al=ctant}{\longrightarrow} &
		- g_2^2 [u_{\al}, u_\beta]^2 = O(u^4) \nonumber\\
	(D_\al \phi)^2
	& \stackrel{u_\al ,\phi =ctant}{\longrightarrow} &
		- g_2^2 [u_{\al}, \phi]^2 = O(\phi^2 u^2)
\label{D=0Operators}
\end{eqnarray}
They give raise to the classical action  ($l=0$) Eq.(\ref{2D}).
But we have also non-classical ($l=1$) contributions to the effective
potential. They correspond to operators arising from two external legs one-loop
graphs $\{b=2, l=1\}$. The one-loop diagram with two external
{\em zero momentum} gluon legs is zero,
as local gauge invariance requires.\footnote{In fact, 2D gauge invariance
prevents the two gluon diagram at zero momentum
to survive to all orders in the loop expansion.
No gluon mass term is allowed by gauge symmetry.} On the contrary the
scalar field is not a 2D gauge field. It can get a mass term through the one
loop diagrams shown in Fig.1, which lead to
\be
 \mu_\phi^2(\epsilon) = \frac{g_2^2 N}{\pi} +
		\frac{g_2^2 N}{2 \pi} \ln(\frac{1}{\epsilon^2})
                                        + O(\epsilon^2 \ln\epsilon)
\label{scalarMass}
\ee
Consequently, for small values of $\epsilon$ dynamics is not given by the
classical action (\ref{ClassicalQCD}) but by the quantum corrected one,

\be
 {\cal S}^{QCD_3} = \int \!d^2x \left\{ \frac{1}{2} f_{\alpha\beta}^2
               +  (D_\alpha \phi)^2 + \mu_\phi^2(\epsilon) \phi^2
                                        + O(\epsilon^2 \ln\epsilon) \right \}
\label{OneLoopQCD}
\ee
which allows us to provide the exact {\it line limit} of $QCD_3$.

The quantum fluctuations generated by the $V$-excitations are able to
provide the classically massless scalar field with a very heavy mass.
Thus the spectrum of $QCD$ on the strip contains just one single massless
field, the 2D gluon. The remaining particles, scalar and excited modes,
are massive.

In $QCD_3$ the decoupling of the heavy sector is caused by quantum effects
in the form of a divergent mass for the scalar particle in the
$\epsilon \arrow 0$ limit,
\be
 \mu_\phi^2(\epsilon) = \frac{g_2^2 N}{\pi} +
		\frac{g_2^2 N}{2 \pi} \ln(\frac{1}{\epsilon^2})
                \stackrel{\epsilon \rightarrow 0}{\rightarrow} \infty
\label{DivergentscalarMass}
\ee
The 2D gauge field is the only massless degree of freedom of the
theory and thus the long range 2D dynamics ($R \gg g_3^{-2}$) is
completely determined by 2D gluon physics.

\s{The spectrum of the effective small $\epsilon$ theory}

The scalar field becomes, like its transversal partners $V$'s,
infinitely heavy  in the completely reduced action ($\epsilon=0$).
In the small $\epsilon$ regime the scalar field $\phi$, although
still very heavy,  is the lightest of all the transversal particles
($\mu_\phi^2 \sim \ln(1/\epsilon^2), m_n^2 \sim 1/\epsilon^2$).
Consequently the appearance of transversal gluons effects takes
place in a much softer way than in the classical reduced action
Eq.(\ref{ClassicalQCD}). However $QCD$ on the strip in the small
$\epsilon$ regime (\ref{OneLoopQCD}) still contains highly non-trivial
gluon physics.

We can take advantage of the fact that the scalar field is very heavy to give a
non-relativistic (NR) treatment to the action (\ref{OneLoopQCD}). Our
interest lies in the calculation of (very massive $\sim \mu_\phi$)
scalar bound states which do not show up in the $\epsilon \rightarrow 0$
limit (pure $QCD_2$).

We are now in an analogous situation to that found in 't Hooft's model. The
heavy scalar field interacts with the 2D gluons as heavy fermions do.
Although fermions and bosons verify different relativistic
equations, they give raise to an identical NR formalism in 2D when masses
are large enough\footnote{In higher dimensions this limit is also the same
with the exception of the term proportional to the spin {\boldmath $S.B$}.
This terms is not present in 2D because of the absence of spin.}. At lowest
order in the inverse mass expansion ($1/\mu_\phi$), the suppressions in the
equation for the bound-state amplitude are the same as those appearing in
leading order in $1/N$ \ci{th74}, namely, absence of sea quark effects,
vertex corrections and non-planar diagram contributions.

The bound state mass equation is nothing but a one dimensional
Schr\"odinger equation for a $\phi_{(1)}\phi_{(2)}$ system interacting
through a color gauge field. Only the zero component of the gauge field
contributes to the potential through the static term $g_2 T^a a_0^a$. Other
gauge field dependent contributions to the potential are suppressed by powers
of $1/\mu_\phi$. In the $a_1=0$ axial gauge, the zero
component of the gauge field is

\begin{eqnarray}
 a_0^a(x_{12}) = - \frac{g_2}{2}T_{(2)}^a |x_{12}|
	& \Rightarrow &
		\hat{V}_{12}=  g_2 T_{(1)}^a a_0^a(x_{12})
		= - \frac{g_2^2}{2} (T_{(1)}^aT_{(2)}^a) |x_{12}|
\label{a0}
\end{eqnarray}
thus the eigenvalue equation for the bound state wave function $\Phi_{12}(x)$
becomes

\be
 -\frac{1}{\mu_R} \frac{d^2}{d x^2} \Phi_{12}(x) +
			\sigma |x| \Phi_{12}(x) = E \Phi_{12}(x)
\label{EigenvalueEquation}
\ee
The string tension $\sigma$ is given in terms of the 2D coupling constant
and the color invariant $C_N$ as  $\sigma = \frac{1}{2} g_2^2 C_N$
\footnote{$C_N$ is the color expectation value
$-\!\!<\!\!(12)|T_{(1)}^a T_{(2)}^a|(12)\!\!>$, where the color state
$|(12)\!\!>$ is \\
a $|T_{tot}=0, \{N^2\!-1\} \otimes \{N^2\!-1\} \!\!>$ state. For $N=2$,
$C_N=3/4$.}.
The mass $\mu_R$ is the renormalized  scalar mass, obtained by
dressing the $\phi$ mass Eq.(\ref{DivergentscalarMass}) by the action of 2D
gluons. The calculation of the
renormalized scalar mass has been carried out in the context of finite
temperature $QCD_3$ under the name of {\em electric mass}, or inverse
Debye screening length  \ci{dh82}, which properly applied to the strip
becomes

\be
 \mu_R^2(\epsilon) = \frac{g_2^2 N}{4 \pi} \ln(\frac{1}{\epsilon^2})
\label{RenormalizedscalarMass}
\ee
The solution of the one dimensional differential equation
(\ref{EigenvalueEquation}) is known. It is given in terms of the Airy
function \ci{dh82,gp78}

\be
  \Phi_r (x) = \mbox{Ai}[(\mu_R \sigma)^{1/3} x - \varepsilon_r]
\label{WaveFunction}
\ee
where $-\varepsilon_r$ is the $r$th zero of Ai or Ai' for odd or even states,
respectively.
The energy of the bound state and thus the mass of the $r$th
glueball is given by

\begin{eqnarray}
 E_r(\epsilon) & =  & 2\mu_R + \varepsilon_r(\frac{\sigma^2}{\mu_R})^{1/3}
					\nonumber \\
               & = & g_2 \sqrt{\frac{2
N}{\pi}}[\ln(\frac{1}{\epsilon^2})]^{1/2}
              + \varepsilon_r g_2
		\left(\frac{C_N^2}{2}\sqrt{\frac{\pi}{2 N}}\right)^{1/3}
                        [\ln(\frac{1}{\epsilon^2})]^{-1/6}
					\nonumber \\
	    &&
\label{GlueballMass}
\end{eqnarray}

The mass of these glueball states grows as $\epsilon$ approaches zero.
They decouple eventually leaving ordinary 2D gluons as the only remnants
of color interaction.

In the small $\epsilon$ regime the lowest glueball spectrum corresponds to
light glueball $\phi\phi$ bound states. But they are certainly not the only
glueball states one can generate from the strip action (\ref{2D}). The
non-zero mode fields $V_n$ are also present. They can likewise yield glueball
particles as massive $V_nV_n^\ast$ bound states. Since $V$ states are heavier
than the scalar particles we expect the lightest $V_nV_n^\ast$ spectrum to be
above the lowest mass scalar glueballs Eq.(\ref{GlueballMass}).

The way to proceed with $V_nV_n^\ast$ bound states is identical to that we
followed for scalar glueballs. First of all, NR approach is even better in this
case than for scalar particles since $\mu_\phi < m_n$ for small $\epsilon$'s.
The only difference lies on the presence of $V$ selfcouplings and of an
additional $V\phi$ interaction in the $V$ action Eq.(\ref{2D}).

As far as the $V$ self-interaction is concerned we can simply neglect it to
lowest order in $\epsilon$. As we saw in Eq.(\ref{Substitution}), in order to
calculate the lowest $\epsilon$ contributions we must perturbate $V$ around its
classical value at $\epsilon=0$. That is, we must expand the non-zero mode
field
$V_n$ around zero

\be
 \hat{V}^n_\al = <\!\!\hat{V}^n_\al\!\!>
		+ \delta\hat{V}^n_\al =  \delta\hat{V}^n_\al
\label{DeltaVn}
\ee
At lowest order in the $\epsilon$ expansion only quadratic terms in $V$ are
relevant in the strip action. Higher power $V$ terms are
suppressed as $\epsilon^2$.

The NR limit of the equation for a $V_nV_n^\ast$  bound state is then
exactly the same as Eq.(\ref{EigenvalueEquation}) except for an extra
interaction term in the potential. This is due to the existence of an
additional
$VV\phi$ vertex supplying an interaction which survives to the NR limit.
However for this piece of the potential is a 2D OSEP (One Scalar Exchange
Potential), it is {\em strongly short range} decaying exponentially with the
scalar mass ($\sim e^{-\mu_\phi|x|}$). As a first approximation (recall
$\mu_\phi$ is very large for small $\epsilon$) we do not consider it and we
evaluate the spectrum taking into account the confining 2D OGEP (One
Gluon Exchange Potential) exclusively. Therefore we can calculate with a
rather good accuracy the $\epsilon$ dependence of the $V$ glueball spectrum
just by copying the result we obtained for scalar glueballs and making the
substitution

\be
 \mu_R \rightarrow m_n
\label{Substitution2}
\ee
That is, the mass of the $r$th glueball state made out of two massive gluons
with $U(1)$ charge $n$ ($V_nV_n^\ast$ state) will be

\begin{eqnarray}
 E_r^{(n)}(\epsilon) & =  & 2m_n +
        \varepsilon_r(\frac{\sigma^2}{m_n})^{1/3}
					\nonumber \\
               & = & g_2 \frac{4 \pi n}{\epsilon} +
                        g_2 \varepsilon_r
                        \left( \frac{C_N^2 \epsilon}{8 \pi n} \right)^{1/3}
					\nonumber \\
	    &&
\label{GlueballMass2}
\end{eqnarray}

The equations (\ref{GlueballMass}) and  (\ref{GlueballMass2}) are valid
for the low energy spectrum. High energy modes (binding energy
$\sim \mu_\phi$) have to be treated in a relativistic framework.
The highest modes (ultrarelativistic ones) can be obtained using the
results of Demerterfi et al. \cite{dk94}\footnote{Notice they calculate
with the {\em massless} adjoint  action (\ref{ClassicalQCD}) which is not
the reduction $QCD_3$ into the line limit.}

\s{Conclusions}

In recent times there have been a growing interest in the study of $QCD_2$
coupled to massless adjoint matter\ci{dk94,sm94}. That is in the study
of the classical reduced action (\ref{ClassicalQCD}). This theory possesses
some appealing properties. It is a supersymmetric theory, it is super-
renormalizable,
it owns a non-trivial glueball spectrum and besides it shows confinement and
asymptotic freedom. It is hoped that, as in 't Hooft's model, one can keep
track of some interesting features of the real theory by studying the action
(\ref{ClassicalQCD}) as a 2D model of $QCD$. This hope would be mainly
based on the dimensional reduction process. In this way the role of the scalar
field would be to give a {\it transversality} component to the theory
\footnote{Recall $\phi$ is nothing but the $A_2$ zero mode.} absent in pure
gauge $QCD_2$.

However, as we have just proved, Eq.(\ref{ClassicalQCD}) does not provide
the right dimensional reduction of $QCD$ in $2+1$ dimensions. The
massless character of the scalar field is a classical feature. Vacuum
fluctuations corresponding to the massive transversal modes renormalize
the scalar mass. This renormalization process is perfectly defined and gives
rise to an $\epsilon$-dependent {\em non-zero} scalar mass
Eq.(\ref{DivergentscalarMass}). Consequently the scalar mass is not an
arbitrary parameter of the theory but a well defined function of the
dimensionless width $\epsilon$.

Another important fact indicated by the small $\epsilon$ expansion action
Eq.(\ref{OneLoopQCD}) is that the {\it line limit} ($\eps \rightarrow 0$)
of gluon $QCD_3$ is gluon
$QCD_2$ . In the framework of an interpolating model of $QCD_3$ like
that given by Eq.(\ref{2D}), this means that $QCD_2$ and the
successive effective theories obtained by increasing the value
of $\epsilon$ are
those which can be connected to real $QCD_3$ in the $\epsilon \gg 1$ regime.
For instance we expect that if there exists some property of the
{\it line limit} ($\epsilon \rightarrow 0$) theory surviving in the
infinite volume limit this will correspond to $QCD_2$ and it will be
respected by all the effective actions interpolating between
($\epsilon \rightarrow 0$) and ($\epsilon \rightarrow \infty$).
The flux generated by the flow of the effective actions with $\epsilon$ is
perfectly defined. For the complexity of the effective 2D theories
representing $QCD_3$ increases rapidly with the value of $\epsilon$
Eq.(\ref{DL}), we come to the conclusion that pure $QCD_2$ and the one
loop action (\ref{OneLoopQCD}) are the simplest 2D models one can relate
to  3D $QCD$\footnote{In terms of 2D gauge degrees of freedom. It is not
excluded an alternative formulation using strings, for example
\ci{gr93}. In any case this approach should be equivalent to that
presented here.}. Certainly there is a limit in the validity of the
perturbative approach. The $\epsilon$ expansion will break down for
large values of $\epsilon$. Nevertheless this flow can be extrapolated
to the non-perturbative region by means of lattice calculations \cite{teper}.
Finite temperature lattice results can be likewise interpreted on the strip
after a suitable euclidean rotation. The interesting finite temperature
operators in the strip framework
are spatial-like, which become time-like ones on the strip.
This is the case of the finite temperature {\em spatial} Wilson loop
corresponding to the ordinary strip Wilson loop.
The analysis of lattice results points out the
existence of a continuous behavior of the
spatial-like Wilson loop over the phase transition point. Moreover it
seems that the spatial-like Wilson loop gets
stabilized already at this point reaching its zero temperature
value. When properly interpreted in the strip language this
fact means that physical properties, as the
glueball spectrum or the string tension, can be continuously extrapolated
over the non-perturbative region
to the three dimensional $\epsilon$ regime \cite{fj94b}.

To end we extend our results to four dimensions in a descriptive fashion.
Consider pure gauge $QCD$ defined on ${\cal R}^2 \otimes
[0,L] \otimes [0,L]$. We perform a double compactification in the
transverse coordinates. The gauge fixing procedure introduces now
two 2D scalar fields instead of one (one per each compactified dimension).
They are massless classically . The existence of heavy non-zero
$U(1) \otimes U(1)$ modes allows the reduction of dynamics to the
scalar sector in the small $L$ regime.
By means of the same compactification techniques used for $QCD$ on the
strip one can construct the effective action \ci{fj94b}. The lowest
order in the $\epsilon$- expansion of the 2D effective action for
$QCD_4$ becomes

\begin{eqnarray}
 \lefteqn{{\cal S}^{QCD_4} =
	 \int \!d^2x \left\{ \frac{1}{2} f_{\alpha\beta}^2 +
        (D_\alpha \phi)^2 + \mu^2(\epsilon) \phi^2  + \right.}
						\nonumber\\
      &&     \left.            +  (D_\alpha \phi')^2 + \mu^{2}(\epsilon)
\phi^{'2}
                + a(\epsilon) \phi \phi'
		- g_2^2 [\phi,\phi']^2
                                        + O(\epsilon^2 \ln\epsilon) \right \}
\label{OneLoopQCD4}
\end{eqnarray}

The integration of the $U(1) \otimes U(1)$ non-zero mode fields is performed
analogously as on the strip. The non-zero mode fields loops
give a very heavy mass ($\mu^2(\eps) \sim \ln(1/\eps)$) to
the scalar sector. They also
give raise to an additional $\phi\phi'$ term.
This crossed term has the effect of breaking the mass degeneracy of the scalar
doublet.
This can be manifestly seen by introducing the field combinations

\begin{eqnarray}
	 \Phi_+ & \equiv  & \frac{1}{\sqrt{2}} (\phi + \phi')
						\nonumber \\
	 \Phi_- & \equiv & \frac{1}{\sqrt{2}} (\phi - \phi')
\label{PhiPhi'}
\end{eqnarray}

which yields to the following effective action in the small $\eps$ regime

\begin{eqnarray}
\lefteqn{ {\cal S}^{QCD_4} =
	\int \!d^2x \left\{ \frac{1}{2} f_{\alpha\beta}^2
         +  (D_\alpha \Phi_+)^2 + \mu_+^2(\epsilon) \Phi_+^2
\right.}
						\nonumber \\
          &&   \left.          +  (D_\alpha \Phi_-)^2
				+ \mu_-^{2}(\epsilon) \Phi_-^{2}
						\right\}
\label{OneLoopQCD42}
\end{eqnarray}
The masses are given by

\begin{eqnarray}
        \mu_+^2(\epsilon) & = & \mu^2(\eps) + \frac{a(\eps)}{2}
			\stackrel{\epsilon \rightarrow 0}{\rightarrow} \infty
							   \nonumber \\
        \mu_-^2(\epsilon) & = & \mu^2(\eps) - \frac{a(\eps)}{2}
			\stackrel{\epsilon \rightarrow 0}{\rightarrow} \infty
\label{NewscalarMasses}
\end{eqnarray}

We see that, like in the strip, scalar particles become extremely heavy in the
small $\epsilon$ regime. They decouple eventually from the 2D gluons in the
$\epsilon \arrow 0$ limit.  The parallelism between the previous action and
the 2D effective action we found for the strip Eq.(\ref{OneLoopQCD}) is
evident. Thus we expect to reproduce {\em qualitatively} the main
features of $QCD$ on the strip at small $\epsilon$'s. The main ingredients are
identical in both theories. Namely, the existence of a very heavy constituent
matter and the presence of a linearly confining interaction supplied by the
exchange of 2D gluons.

\s{Acknowledgments}
The authors are specially grateful to Uwe-Jens Wiese, who has taken part
in the first stage of the present work. We also wish to thank V. Vento
for extensive discussions and his strong support. One of us (A.F.) has
benefitted from iluminating discussions with P. Hasenfratz, P. Minkowski,
J. Polonyi, B. M\"uller, J.I. Latorre and J. Tar\'on. A.J. thanks P.
Gosdzinsky for fruitful discussions.

\newpage
\begin{center}
\large{Figure Captions}
\end{center}
Fig.1 : Diagrams contributing to the scalar mass in lowest order in
$\epsilon$. Solid lines represent adjoint scalar fields $\phi$.
Double wavy lines
represent heavy vector modes $V_\al^n$.
\end{document}